\documentclass[pre,aps,twocolumn,nofootinbib]{revtex4-1}
\usepackage[latin1]{inputenc}
\usepackage[english]{babel}
\usepackage{graphicx}
\usepackage{color}
\usepackage{amsmath}
\usepackage{amssymb}
\usepackage{epstopdf}

\begin{document}

\title{\bf Effective geometries and generalized uncertainty principle corrections to the Bekenstein--Hawking entropy}
%
%
\author{Ernesto Contreras}
\affiliation{Centro de F\'{\i}sica Te\'orica y Computacional,
Facultad de Ciencias, Universidad Central de Venezuela,
AP 47270, Caracas 1041-A, Venezuela.}
\author{Fabi\'an D. Villalba}
\affiliation{Departamento de F\'{\i}sica,
Universidad de los Andes, Apartado A\'ereo {\it 4976}, Bogot\'a, Distrito Capital, Colombia}
\author{Pedro Bargue\~no}
\email{p.bargueno@uniandes.edu.co}
\affiliation{Departamento de F\'{\i}sica,
Universidad de los Andes, Apartado A\'ereo {\it 4976}, Bogot\'a, Distrito Capital, Colombia}
\begin{abstract}
In this work we construct several black hole metrics which are consistent with the generalized uncertainty principle 
logarithmic correction to the Bekenstein--Hawking entropy
formula. After preserving the event horizon at the usual position, a singularity at the Planck scale is found. Finally, these geometries are shown to
be realized by certain model of nonlinear electrodynamics, which resembles previously studied regular black hole solutions. 
\end{abstract}

\maketitle

\section{Introduction}
Black hole (BH) entropy can be considered as the paradigmatic quantum gravitational effect {\it par excellence} one can think of.
After the initial findings by Bekenstein \cite{Bekenstein1972,Bekenstein1973,Bekenstein1974}, Hawking realized 
\cite{Hawking1974,Hawking1975},
within the framework of quantum field theory in curved backgrounds, that BHs radiate. The entropy of a Schwarzschild
BH is given by the Bekenstein--Hawking relation
\begin{equation}
S= \frac{\mathcal{A}}{4l_{p}^{2}},
\end{equation}
where $\mathcal{A}$ is the area of the BH horizon and $l_{p}=\sqrt{\frac{G\hbar}{c^{3}}}$ is the Planck length.

In the quest for a complete theory of quantum gravity (QG), several approaches to it have predicted 
particular forms for the QG--corrected BH entropy
\cite{Kaul2000,Medved2004,Camelia2004,Meissner2004,Das2002,Domagala2004,Chatterjee2004,Akbar2004,Myung2004,Chatterjee2005}.

For example, starting from the quadratic generalized uncertainty principle (GUP) 
\cite{Medved2004}
\footnote{The GUP gives rise to a minimal length scale 
which is thought to be a essential ingredient of any quantum gravitational theory \cite{Sabine2013}.},
whose effects 
can be implemented both in classical and quantum systems by defining deformed
commutation relations by means of \cite{Dass2008}
\begin{equation}
x_{i}=x_{0i}\; ;  \; p_{i}=p_{0i}\,\left(1+ 2 \alpha^{2} p_{0}^{2}\right),
\end{equation}
where $\left[x_{0i},p_{0j}\right]=i \hbar \delta_{ij}$ and $p_{0}^{2}=\sum_{j=1}^{3}p_{0j}p_{0j}$ and
$\alpha = \nobreak \alpha_{0}/m_{p}c$, being $\alpha_{0}$ a dimensionless constant, it is shown that  
the corrected BH entropy can be written as
\begin{equation}
\label{eqfirst}
S=\frac{\mathcal{A}}{4 l_{p}^{2}}-\frac{\pi\alpha^{2}}{4}\ln \left(\frac{\mathcal{A}}{4 l_{p}^{2}} \right)
+\sum\limits_{n=1}^{\infty}c_{n}\left(\frac{\mathcal{A}}{4l_{p}^{2}}\right)^{-n}+const
\end{equation}
where $c_{n}=\alpha^{2(n+1)}$.
 We will take $\hbar=c=G=1$. Therefore, within this choice, $m_{p}=l_{p}=1$ and $\alpha=\alpha_{0}$. 
It is noteworthy that the Loop Quantum Gravity (LQG) prediction is obtained
by considering $\alpha=\sqrt{2/\pi}$. 

Recently, Scardigli and Casadio
\cite{Scardigli2015} proposed a deformed spherically symmetric and static Schwarzschild metric using the ansatz 
\begin{equation}
f(r)=1-\frac{2M}{r}+\epsilon \frac{M^2}{r^2}
\end{equation}
for the time--time component of the metric to reproduce the modified Hawking temperature as a consequence of the GUP. 
As pointed out very recently by 
A. F. Ali, M. M. Kahlil and E. C. Vagenas \cite{Vagenas2015}, this
ansatz implies a different position for the event horizon, contrary to many arguments based on the
GUP \cite{Scardigli1999,Maggiore1993,Camelia2006}. In
Ref. \cite{Vagenas2015}, the authors extend the class of 
metrics which give place to the GUP--corrected Hawking temperature by assuming a functional
dependence of the form
\begin{equation}
f(r)=\left(1-\frac{2M}{r}\right)\left(1+\eta \left(\frac{2M}{r} \right)^n \right),
\end{equation}
where $\eta\ll1$ is a constant and $n\ge0$ is an integer. 
After comparing the modified Newton law with the Randall--Sundrum II model \cite{RS1999}, the 
authors of \cite{Vagenas2015} conclude that the most likely value for $n$ is $n=2$.

In this work we will look for a reinterpretation of the first and second terms of the RHS of Eq. (\ref{eqfirst}) in a semi-classical way.
Specifically, we will look for
spherically symmetric and static geometries whose surface gravity at the horizon leads to the Bekenstein--Hawking {\it plus} the logarithmic
correction Eq. (\ref{eqfirst}). Thus,
our approach tries to incorporate
some GUP--related quantum gravitational effects in terms of geometries which satisfy Einstein's equations.  

In this sense, this work constitutes a semiclassical approach to the BH entropy. Interestingly, there have been also other works of
semiclassical nature which try to solve the BH singularity problem by introducing modifications of the spherically 
symmetric Hamiltonian constraint in terms
of holonomies (see, for example, \cite{Modesto2006,Modesto2008,Modesto2010} and references therein). Moreover, it
is noteworthy that, although the methods employed by
the author of Refs. \cite{Modesto2006,Modesto2008,Modesto2010} are very different from ours, there are resemblances 
between some conclusions reached by these two approaches, as will be commented along the manuscript.

The paper is organized as follows. In section II, we briefly present how to derive BH geometries which
include GUP effects on the entropy the Schwarzschild BH and the main properties of some of these geometries are 
analized in terms of certain 
deformations of the Schwarzschild solution. In section III,
we will interpret the previous geometries in terms of gravity coupled to non--linear electrodynamics showing that our 
findings indicate the presence of a non--linear
Reissner--Nordstr\"om (RN) BH. Finally, in section IV, a brief summary of the obtained results is given.
%

\section{Semi-classical metrics for GUP--corrected black hole entropy}

The main idea is to obtain a spherically symmetric and static exact solution of Einstein's equations 
such that its corresponding entropy, computed from the semi-classical gravity at the horizon, 
incorporates the GUP logarithmic correction. As commented
along the Introduction, we demand that the location of the horizon of this proposed 
solution coincides with that of the Schwarzschild case, in agreement with
many arguments based on the GUP \cite{Scardigli1999,Maggiore1993,Camelia2006}. Therefore, for a
Schwarzschild--deformed metric of the form
\begin{equation}
ds^{2}=-f(r)dt^{2}+f(r)^{-1}dr^{2}+r^{2}d\Omega^{2},
\end{equation}
these requirements read $\beta=\frac{2\pi}{\kappa}$, where $\kappa=\frac{f'(r_{H})}{2}$ 
is the surface gravity at the horizon $r_{H}=2M$ and $\beta$ is the inverse temperature of the BH. The prime
denotes differentiation with respect to the radial variable.

If the ansatz function is taken to be of the form,
\begin{eqnarray}
\label{ansatz}
f(r)&=&\left(1-\frac{2M}{r}\right) g(r) \\
&& \nonumber
\end{eqnarray}
after using the second law as $dS=\beta dM$, the following deformed--inverse temperature is obtained 
\footnote{The Planck mass has been incorporated in order 
to have a dimensionally correct expression. In subsequent expressions, 
also the Planck length (mass) will be sometimes incorporated to clarify the discussion.}
\begin{equation}
\label{morebeta}
\beta=8\pi M\left[1-\left(\frac{\alpha m_{p}}{4 M}\right)^2 \right].
\end{equation}

Therefore, from the standard relations between $\beta$ and $\kappa$, $g(r_{H})$ reads
\begin{center}
\begin{equation}
g(r_{H})=\left[1-\left(\frac{\alpha l_{p}}{2 r_{H}}\right)^2 \right]^{-1}.
\end{equation}
\end{center}

Let us note that one possible choice for $g(r)$ that satisfies the previous requirements gives place to a family of
functions given by
\begin{equation}
g(r)=\left(1-\frac{ \alpha^2 l_{p}^{2} (2M)^n}{4 r^{n+2}}\right)^{-1}.
\end{equation}

After a long but straightforward calculation, the algebraic curvature invariants reveal
that there is an intrinsic singularity at $r_{s}=\alpha\, l_{p}/2$ when $n=0$. 
This means that the breakdown of classical general relativity occurs within a region whose length
scale is the Planck length, as one should expect (we remind the  reader that $\alpha$ is a dimensionless constant of order unity).
It is interesting to note that the $S^{2}$ sphere of the LQG BH case \cite{Modesto2010} bounces on the minimum area of LQG and
the singularity dissapears. In our case, the singularity is still present, but this time near the Planck length.
In the case of $n\ne0$  solutions there is also an intrinsic singularity at
$r_{s}=\left[\alpha^2 l_{p}^{2}(2M)^{n}/4\right]^{1/(n+2)}$. In these cases, the singular region depends not only on $l_{p}$ but also on the mass $M$ of the
gravitating object. Therefore, we do not consider them as physically relevant as only the Planck scale is expected to be linked to the scale where 
quantum gravitational effects become dominant.

Thus, returning to the case $n=0$, let us note that, considering the asymptotic behavior of $f(r)$ we obtain 
\begin{equation}
f(r)\longrightarrow 1-\frac{2M}{r}+\frac{\alpha^{2}}{4r^{2}}.
\end{equation}
Therefore, within this limit, the geometry can be interpreted as that of a deformed RN BH with an electric charge $q$ such that $\alpha=2 q$
(note that some similarities between the LQG and the RN BHs were pointed out in Refs. \cite{Modesto2006,Modesto2008} concerning
mainly the causal structure of these spacetimes).

At this point, let us summarize our main findings:

\begin{itemize}
\item We have shown that the entropy associated to the deformed Schwarzschild metric corresponds to that of the first two terms
of the RHS of Eq. (\ref{eqfirst}).
\item This deformed metric has an intrinsic singularity located at the Planck scale.
\item At infinity, the metric behaves as that of a charged and static BH with $\alpha=2 q$.
\end{itemize}

Then, the next step is to look for a possible interpretation of the deformed metric, which we write as
\begin{widetext}
\begin{equation}
\label{our}
ds^{2}=-\left(1-\frac{2M}{r}\right)\left(1-\frac{q^2}{r^2}\right)^{-1}dt^{2}+
\left(1-\frac{2M}{r}\right)^{-1}\left(1-\frac{q^2}{r^2}\right)dr^{2}+r^{2}d\Omega^{2}.
\end{equation}
\end{widetext}
Given the previous RN--like interpretation at spatial infinity,
it seems plausible to impose the geometry to be a solution to the Einstein--Maxwell system, when certain non--linear electrodynamics is invoked.

\section{Coupling gravity to non--linear electrodynamics}

By Israel's theorem, the only electrovacuum static and spherically symmetric solution of the Einstein--Maxwell system is 
the RN one \cite{Israel1967}. Therefore,
Eq. (\ref{our}) can not be a solution of this coupled system. However, our deformed Schwarzschild metric will appear when coupling 
gravity to a certain
non--linear electrodynamics (NLED) theory. The importance of these theories is twofold: first, quantum corrections to Maxwell theory can be 
described by means of non--linear effective Lagrangians that define NLEDs as, for example,
the Euler--Heisenberg Lagrangian \cite{HE,Sch}, which can be effectively described using Born--Infeld (BI) theory \cite{BI}. 
Second, it is well known that in case of dealing with open bosonic strings, the resulting tree--level
effective Lagrangian is shown to coincide with the BI Lagrangian \cite{PLB1985,NPB1997}. Apart from gravitational BI solutions
\cite{Garcia1984,Breton2003}, an exact regular BH geometry in the presence of NLED was obtained in \cite{Ayon1998} and further discussed in
\cite{Baldo2000,Bronni2001}. In addition, the same type of solutions with Lagrangian densities that are powers of Maxwell's Lagrangian were analyzed in 
\cite{Hassaine2008}. Recently, a wide family of regular BHs satisfying the weak energy condition has been presented \cite{Balart2014a,Balart2014b}.

Let us consider the following energy--momentum tensor for NLED:
\begin{equation}
\label{nlT}
T^{\mu\nu}=-\frac{1}{4\pi}\left[\mathcal{L}(F)g^{\mu\nu}+\mathcal{L}_{F}F^{\mu}_{\;\;\rho}F^{\rho \nu} \right],
\end{equation}
where $\mathcal{L}$ is the corresponding Lagrangian, $F=-\frac{1}{4}F_{\mu\nu}F^{\mu\nu}$ and $\mathcal{L}_{F}=\frac{d\mathcal{L}}{dF}$. 

Assuming spherically symmetric and static electrovacuum solutions and taking only a radial electric field as the source, that is,
\begin{equation}
\label{eqmaxwell}
F_{\mu \nu}=E(r)\left(\delta^{r}_{\mu}\delta^{t}_{\nu}-\delta^{r}_{\nu}\delta^{t}_{\mu} \right),
\end{equation}
Maxwell equations read
\begin{equation}
\label{motion}
\nabla_{\mu}\left(F^{\mu\nu}\mathcal{L}_{F}\right)=0.
\end{equation}
Thus, from \eqref{eqmaxwell} and \eqref{motion} one can obtain an explicit expression for the electric field
\begin{equation}
\label{max}
E(r)=-\frac{q}{r^{2}}(\mathcal{L}_{F})^{-1}.
\end{equation}
After some algebraic computations, the electric field is shown to be given by
\begin{equation}\label{electricfull}
E(r)=\frac{m'(r)}{q}-\frac{r}{2q}m''(r),
\end{equation}
where the mass function $m(r)$ is such that $f(r)=1-2m(r)/r$. 

In our case, using Eq. (\ref{our}), $m(r)$ results to be
\begin{equation}
m(r)=\frac{r \left(q^2-2 M r\right)}{2 \left(q^2-r^2\right)}
\end{equation}
and the corresponding electric field is given by
\begin{eqnarray}
E(r)&=&\frac{q \left(q^4+2 (5 M-r) r^3-q^2 r (2 M+3 r)\right)}{2 \left(q^2-r^2\right)^3} \nonumber \\
&=&\frac{q}{r^2}+\mathcal{O}\left[r\right]^{-3}.
\end{eqnarray}
Therefore, Eq. (\ref{our}) behaves as a RN BH at infinity, which supports our description 
in terms of NLED.

The underlying NLED theory can be obtained using the $P$ framework \cite{Salazar1987}, which is somehow dual to the $F$ framework. 
One introduces the tensor $P_{\mu\nu}=\mathcal{L}_{F}F_{\mu\nu}$ together with its invariant $P=-\frac{1}{4}P_{\mu\nu}P^{\mu\nu}$ 
and considers the 
Hamiltonian--like quantity
\begin{equation}
\label{H}
\mathcal{H}=2 F \mathcal{L}_{F} -\mathcal{L}
\end{equation}
as a function of $P$. This quantity $\mathcal{H}(P)$ specifies the theory. The Lagrangian can be written as a function of $P$
as
\begin{equation}
\label{L}
\mathcal{L}=2 P \frac{d\mathcal{H}}{d{P}}-\mathcal{H}. 
\end{equation}
Finally, by reformulating the energy--momentum tensor in terms of $P$, $\mathcal{H}(P)$ is shown to be given by \cite{Bronni2001}
\begin{equation}
\mathcal{H}(P)=-\frac{1}{r^2}\frac{d m(r)}{dr}.
\end{equation}

In our case, and considering only the Hamiltonian function for simplicity, the NLED can be shown to be given by
\begin{equation}
\label{HEP}
\mathcal{H}(P)=-\frac{P \left(1+\sqrt{2 P q^2}-\frac{2^{5/4} P^{1/4} \sqrt{q}}{s}\right)}{1+2 P q^2-\sqrt{8 P q^2}}.
\end{equation}
where the parameter $s=q/2M$ has been introduced to facilitate comparison with \cite{Ayon1998} (see the following discussion).

After Taylor expanding Eq. (\ref{HEP}) we get
\begin{widetext}
\begin{equation}
\label{Taylor}
\mathcal{H}(P)=-P+\frac{ 2^{5/4} \sqrt{q} P^{5/4}}{s}-3 \sqrt{2 q^2} P^{3/2}+\frac{2^{11/4} q^{3/2} P^{7/4}}{s}-10\, q^2 P^2+
\mathcal{O}[P]^{9/4}.
\end{equation}
\end{widetext}
A couple of comments are in order here. First of all, let us note that Maxwell's theory, $\mathcal{H}(P)=-P$, is
recovered for small fields. In
addition, a quadratic BI--like term appears (fifth term in the RHS of Eq. (\ref{Taylor})). This quadratic term is easy 
to interpret in light
of the cutoff field which is an essential ingredient of BI--theory. Therefore, the difficulty of interpreting this NLED 
theory can
be ascribed to the other terms which appear in the RHS of Eq. (\ref{Taylor}). In spite of this, let us note that similar 
terms have appeared since 
the discovery of the first exact regular BH solution by Ay\'on--Beato and Garc\'{\i}a \cite{Ayon1998}. In fact, the
Hamiltonian function presented in
Ref. \cite{Ayon1998} can be expanded for weak fields to give
\begin{widetext}
\begin{equation}
\mathcal{H}_{AB}(P)=-P+\frac{3\cdot  2^{1/4} \sqrt{q} P^{5/4}}{s}-6 \sqrt{2 q^2} P^{3/2}+\frac{15 q^{3/2} P^{7/4}}{2^{1/4} s}-30 q^2 P^2 +\mathcal{O}[P]^{9/4},
\end{equation}
\end{widetext}
which except for some constants coincide with our Eq. (\ref{Taylor}).

We note that, although there are some similarities between the solution of Ay\'on--Beato and Garc\'{\i}a and
our Eq. (\ref{our}), there are some essential 
differences between them, mainly concerning the
weak energy condition (WEC), which states that the local energy density cannot be
negative for all observers. For the metrics here considered, the WEC can
be stated as
\begin{eqnarray}
\frac{1}{r^2}\frac{dm(r)}{dr}&\ge& 0 \nonumber \\
\frac{2}{r}\frac{dm(r)}{dr}-\frac{d^2m(r)}{dr^2} &\ge 0&.
\end{eqnarray}
Therefore, it is easy to see that Eq. (\ref{our}) violates the WEC. However,
as can be shown by direct calculations, this violation is proportional to $\alpha^2$.
Moreover, our solution can be considered perturbative in the following sense. We have considered
corrections of order $\alpha^2$ to the geometry which are compatible with a logarithmic correction to the BH
entropy as predicted by a quadratic GUP.
Furthermore, we have shown that the RHS of Einstein equations can be interpreted as some kind of NLED. The point is to
note that also this NLED depends
perturbatively on $\alpha^2$ due to the fact that $\alpha = 2 q$. Therefore, we are in a situation similar to that 
of Ref. \cite{Vagenas2015} (where also
the WEC is violated at order $\alpha^2$) but in our case with a completely specified matter content which gives place 
to the required entropy corrections.

\section{Conclusions}
Although the search for a quantum theory of gravity is still under progress, some results about the behavior of the space--time at the
Planck scale can be adscribed to the existence of a minimum length \cite{Sabine2013}, which could be realized by a GUP which, among 
other implications,
gives place to a logarithmic correction to the Bekenstein--Hawking black hole entropy. In this work we have proposed a deformation 
of the Schwarschild 
metric which gives place to this logarithmic correction in a semiclassical way. 
This deformation preserves the location of the event horizon (as required by the GUP approach) and predicts the existence of a 
singularity
at the Planck scale. Moreover, we have shown that this geometry is realized when gravity is coupled to a nonlinear electrodynamics 
model, obtaining an
exact solution which has some resemblances with other well known regular black hole solutions. Although the weak energy condition 
is violated 
at second order in the GUP parameter (also reported in a recent work \cite{Vagenas2015}), it would be interesting to investigate 
whether or not is possible 
to obtain effective geometries with reproduce the logarithmic correction without violating this energy condition. 
We hope to report on this in a future work. \\
\\
P. B. acknowledges support from the Faculty of Science and Vicerrector\'{\i}a de Investigaciones of
Universidad de los Andes, Bogot\'a, Colombia.

\end{document}